# The Halo and Rings of the Planetary Nebula NGC 40 in the Mid-Infrared

G. Ramos-Larios[1], J.P. Phillips[1], L.C. Cuesta[2],

[1]Instituto de Astronomía y Meteorología, Av. Vallarta No. 2602, Col. Arcos Vallarta, C.P. 44130 Guadalajara, Jalisco, México   e-mails : gerardo@astro.iam.udg.mx; jpp@astro.iam.udg.mx

[2]Centro de Astrobiología,  Instituto Nacional de Técnica Aeroespacial, Carretera de Ajalvir, km 4, 28850 Torrejón de Ardoz, Madrid, Spain
e-mail : cuestacl@inta.es

**Abstract**

We present imaging and spectroscopy of NGC 40 acquired using the Spitzer Space Telescope (Spitzer), and the Infrared Space observatory (ISO). These are used to investigate the nature of emission from the central nebular shell, from the nebular halo, and from the associated circumnebular rings. It is pointed out that a variety of mechanisms may contribute to the mid-infrared (MIR) fluxes, and there is evidence for a cool dust continuum, strong ionic transitions, and appreciable emission by polycyclic aromatic hydrocarbons (PAHs). Prior observations at shorter wavelengths also indicate the presence of warmer grains, and the possible contribution of $H_2$ transitions. Two of these components (warm dust with $T_{GR}$ ~ 500-620 K and PAH emission bands) appear capable, in combination, of explaining the infrared colours of the rings and halo, although the flux ratios are also consistent with shock-excited $H_2$ v = 0-0 emission. It is noted that whilst the surface brightness of the rings is greater in the longer wave (5.8 and 8.0 $\mu$m) photometric channels, their fractional fluxes (when compared to the halo) are greater at 3.6 and 4.5 $\mu$m; a trend which is similar to those observed in other planetary nebulae. It is also apparent that the relative intensities of the rings are greater than is observed for the H$\alpha$+[NII] transitions. It is suggested that an apparent jet-like structure to the NE of the halo represents one of the many emission spokes that permeate the shell – and which are observed for the first time in these MIR results. The spokes are likely to be caused by the percolation of UV photons through a clumpy interior shell, whilst the jet-like feature is enhanced due to locally elevated electron densities; a result of interaction between NGC 40 and the interstellar medium. It is finally noted that the presence of the PAH, 21 $\mu$m and 30 $\mu$m spectral features testifies to appreciable C/O ratios within the main nebular shell. Such a result is consistent with abundance determinations using collisionally excited lines, but not with those determined using optical recombination lines.

**Key Words:** (ISM:) planetary nebulae: individual: NGC 40 --- ISM: jets and outflows --- infrared: ISM --- ISM: lines and bands

## 1. Introduction

NGC 40 has several unusual physical and morphological characteristics which place it apart from the majority of Galactic planetary nebulae (PNe). Thus, it appears that the central star has a WC8 spectrum (Smith & Aller 1969; Smith 1973; Mendez & Niemela 1982; Crowther et al. 1998) high rates of mass-loss (of the order of $10^{-6}$-$10^{-5}$ $M_{\odot}$ yr$^{-1}$; e.g. Bianchi 1992; Bianchi & Grewing 1987; Marcolino et al. 2007); very low fractional abundances of H, and high abundances of C ($\beta$(H) < 2%; $\beta$(C) ~ 51% (Marcolino et al. 2007; Leuenhagen et al. 1996)), and evidence for spectral variability on a scale of hours (Balick et al. 1996; Grosdidier, Acker & Moffat 2001) - a property which is likely to arise from the ejection of a clumpy wind with velocity ~1-1.8 $10^3$ km s$^{-1}$ (Bianchi & Grewing 1987; Leuenhagen et al. 1996; Feibelman 1999; Heap 1986). Modelling of the central star spectrum implies temperatures ~ 73-90 kK (Bianchi & Grewing 1987; Leuenhagen et al. 1996; Marcolino et al. 2007).

The nebular shell, on the other hand, has an extremely low level of excitation (excitation class 2), in which the [OII] $\lambda$3727 doublet represents the strongest component of emission; the normally dominant [OIII] $\lambda\lambda$4959+5007 transitions are in this case relatively weak (e.g. Aller et al. 1972). Corresponding estimates of the Zanstra and Stoy temperatures are similarly low, and imply values ~30-38 kK (Koppen & Tarafdar 1978; Harman & Seaton 1966; Preite-Martinez & Pottasch 1983; Pottasch et al. 2003); less than half of that estimated from the central star (CS) continuum.

Various suggestions have been made to explain these disparities, including the possibility that higher energy UV photons are being absorbed by a CII "curtain" (Bianchi & Grewing 1987). It has also been suggested that the higher temperature determination by Bianchi & Grewing may be unreliable because of the values of $E_{B-V}$ which they employ, and the insensitivity of their procedure to estimates of $T_{EFF}$ (Pottasch et al. 2003). Leuenhagen et al. (1996), on the other hand, note that the Zanstra & Stoy temperatures are themselves open to uncertainty, and just as likely to lead to unreliable results.

It is most likely, however, that both of these results are qualitatively correct – that shell Zanstra & Stoy temperatures are low, whilst central star effective temperatures are much higher. It has been noted for instance that much of the CS radiation is likely to be blocked by the stellar wind (Leuenhagen et al. 1996, Morcolina et al. 2007), and that this would lead to strong reductions in the levels of emergent flux – and lower excitation of the nebular shell.

Such anomalies in the Zanstra and CS temperatures also have their counterparts in the analysis of elemental abundances. Thus, it has been noted that abundances determined using optical recombination lines (ORLs) are systematically higher that those determined using collisionaly excited lines (CELs) – a disparity which leads to factors of ~17 difference for the O abundances in this source (Liu et al. 2004). It was originally suggested that such a disparity could be explained where there are appreciable variations in electron temperature $T_e$, although this doesn't explain the persistence of these differences where temperature insensitive transitions are employed (Pottasch et al. 2003). The difference may also arise where OII recombination lines probe H-poor (oxygen rich) plasmas, which take the form of denser clumps or filaments (Liu et al. 2000; Tsamis et al. 2004). An interesting aspect of this difference, at least for the case of NGC 40, is that the ORL and CEL abundances also predict differing values for C/O – C/O is greater than unity for the CEL results, and < 1 for the ORL results. Given that the MIR spectrum of the nebula also depends upon the relative abundances of C and O, we shall see that the Spitzer and ISO results permit us to test these two conflicting abundance procedures. Where C/O is greater than unity, then one might anticipate that PAH features, SiC bands, and/or other C-rich indicators would be present, whilst oxygen rich shells might be expected to show evidence for amorphous and/or crystalline silicate emission.

Finally, we note that the structure of the envelope is also relatively unusual, and shows an elliptical higher excitation ([OIII]) spatio-kinematic structure (Louise 1981; Sabbadin et al. 2000; Sabbadin & Hamzaoglu 1982; Meaburn et al. 1996). The lower excitation [NII] lines, on the other hand, are more spatially extended, and suggest a cylindrical or bipolar outflow oriented along the major axis of the shell (Sabbadin et al. 2000; Meaburn et al. 1996) – a structure which is also consistent with the radio mapping of George et al. (1974). This interior envelope is surrounded by a fainter, more spherical halo with diameter

>90 arcsec; a structure whose kinematics have been investigated by Meaburn et al. (1996).

There is also evidence that the inner nebular components (the core and inner halo) are surrounded by a much weaker filamentary halo extending over ~4.4 arcmin, most clearly to be seen in the imaging of Balick et al. (1992). Martin et al. (2002) have interpreted the presence of bright filaments to the NE as evidence for interaction of NGC 40 with the interstellar medium (ISM), and this may also explain the asymmetric distribution of fainter emission noted by Balick et al. (1992), where it is found that the halo extends preferentially towards the south. Closer into the nucleus, on the other hand, in the inner portions of the AGB halo, it has also been noted that there is evidence for several narrow annular rings (Corradi et al. 2004); features which have also been observed in ~9 other PNe (see e.g. Corradi et al. 2004; Phillips et al. 2009; Phillips & Ramos-Larios 2010), but whose origins remain still to be determined (see e.g. Corradi et al. 2004; Phillips et al. 2009 & references therein).

We shall present ISO and Spitzer spectroscopy of the inner regions of NGC 40, noting the abundance of carbon-rich features in the MIR. It will be suggested that this is consistent with abundances derived using CELs, but conflicts with those determined using ORLs. We shall also provide mapping of the source in the MIR, and undertake an analysis of emission deriving from the halo and rings. It will be suggested that much of the halo emission may arise from combination of warm dust continua ($T_{GR}$ ~ 500-620 K) and PAH emission bands, although it may also derive from shock excited molecular hydrogen.

## 2. Observations

We shall be making use of imaging and spectroscopy of NGC 40 deriving from Spitzer programs 93 (Survey of PAH Emission, 10-19.5 μm) and 40115 (Dual Dust chemistry in Wolf-Rayet Planetary Nebulae), as well as spectroscopy deriving from the Infrared Space observatory (ISO)[1].

---

[1] *Based on observations with ISO, an ESA project with instruments funded by ESA Member States (especially the PI countries: France, Germany, the Netherlands and the United Kingdom) and with the participation of ISAS and NASA.*

The Spitzer imaging of NGC 40 with the Infrared Array Camera (IRAC; Fazio et al. 2004) took place on 17/10/2007, and the results have been processed as described in the IRAC data handbook (available at http://ssc.spitzer.caltech.edu/irac/dh/iracdata_handbook_3.0.pdf). The resulting post-Basic Calibrated Data (post-BCD) are relatively free from artefacts; well calibrated in units of MJy sr$^{-1}$; and have reasonably flat emission backgrounds. The observations employed filters having isophotal wavelengths (and bandwidths $\Delta\lambda$) of 3.550 μm ($\Delta\lambda$ = 0.75 μm), 4.493 μm ($\Delta\lambda$ = 1.9015 μm), 5.731 μm ($\Delta\lambda$ = 1.425 μm) and 7.872 μm ($\Delta\lambda$ = 2.905 μm). The normal spatial resolution for this instrument varies between ~1.7 and ~2 arcsec (Fazio et al. 2004), and is reasonably similar in all of the bands, although there is a stronger diffraction halo at 8 μm than in the other IRAC bands. This leads to differences between the point source functions (PSFs) at ~0.1 peak flux.

We have used these data to produce colour coded combined images of the source in three IRAC bands, where 3.6 μm is represented as green, 4.5 μm as red, and 5.8 μm is indicated as blue. These results have also been processed using unsharp masking techniques, whereby a blurred or "unsharp" positive of the original image is combined with the negative. This leads to a significant enhancement of higher spatial frequency components, and an apparent "sharpening" of the image (see e.g. Levi 1974). The 8.0 μm images were affected by saturation of the bright central star, and a broad emission band across the centre of the source. The use of this data has therefore been confined to investigating flux ratios and profiles.

Profiles through the source were obtained after correcting for the effects of background emission; a component which is present in all of the bands, but appears to be particularly strong at 5.8 and 8.0 μm. This involved the removal of large background offsets, as well as much smaller linear gradients. We have also obtained contour mapping of the 8.0μm/4.5μm, 5.8μm/4.5μm and 3.6μm/4.5μm flux ratios. This was undertaken by estimating the levels of background emission, removing these from the 3.6, 4.5, 5.8, and 8.0 μm images, and setting values at < 3$\sigma_{rms}$ noise levels to zero. The maps were then ratioed on a pixel-by-pixel basis, and the results contoured using standard IRAF programs.

Some care must be taken in interpreting the latter results, however, since scattering within the IRAC camera can lead to errors in relative intensities. The flux corrections described in Table 5.7 of the IRAC handbook suggests maximum changes of ~0.944 at 3.6 μm, 0.937 at 4.5 μm, 0.772 at 5.8 μm and 0.737 at 8.0 μm, although the precise values of these corrections also depends on the distribution of surface brightness within the source. We have, in the face of these problems, chosen to leave the flux ratio mapping unchanged. The maximum correction factors for the 8.0μm/4.5μm and 5.8μm/4.5μm ratios are likely to be > 0.8, but less than unity, and ignoring this correction has little effect upon our interpretation of the results.

Finally, we have provided contour mapping of NGC 40 in which the lowest contours are set at 3σ background noise levels or greater. The intrinsic surface brightness $E_n$ is given by $E_n = A10^{(n-1)C} - B$ MJy sr$^{-1}$, where n is the contour level ($n = 1$ corresponds to the lowest (i.e. the outermost) level), and B is background. Details of these parameters are provided in the caption to the figure.

Spectroscopy of NGC 40 was obtained through Spitzer Program 93 (Survey of PAH Emission, 10-19.5), in which the central portions of the source (α(2000.0) = 00h 13m 00.82s; δ = +72° 31' 18.3") were observed using the Short-High (SH) module of the Infrared Spectrograph (IRS; Houck et al. 2004). The slit dimensions were 11.3x4.7 arcsec$^2$, the spectral resolution of order ~600, and the spectrum extended between 9.9 and 19.6 μm. Similarly, we make use of spectroscopy deriving from the ISO, a 60 cm f/15 telescope which was launched on 17/11/1995. This carried a Short Wave Spectrometer (SWS; de Grauuw et al. 1999) operating between 2.4 and 45 μm, and had aperture sizes varying from 14x20 arcsec$^2$ for λ = 2.38→12.1 μm, 14x27 arcsec$^2$ between λ = 12→29 μm, and 20x22 arcsec$^2$ between 29 and 45.3 μm. The absence of appreciable discontinuities between these differing sectors of the SWS spectrum (Fig. 1; i.e. at the wavelengths λ = 12.1 μm, 29 μm and 45.3 μm corresponding to changes of aperture size) suggests that much of the shorter wave emission (λ < 30 μm) is contained within the aperture areas – although it's possible that the rapid rise of the 30 μm feature close to 29 μm is influenced by the 16% change in aperture size. A comparison between the fluxes of the Spitzer and ISO spectra suggests a flux difference of ~5 which is close to the relative areas of the SWS/IRS apertures.

By contrast, the ISO Long Wave Spectrometer (LWS; Clegg et al. 1999) operated between 45 and 198.8 μm. The beam was elliptical as a consequence of the secondary mirror supports, and the inclination of the LSW mirror M2 (Clegg et al. 1996). The FWHM also appears to have varied very slightly with wavelength, although its mean value was of the order of ~70 arcsec (see e.g. Polehampton 2002). We have reduced the LWS spectrum by a factor 2.8 to match the SWS spectra close to 44 μm (see our further discussion in Sect. 3).

The SWS spectrum of the central parts of the source ($\alpha(2000.0)$ = 00h 13m 01.14 s; $\delta$ = +72° 31' 18.9") was taken through observation TDT30003803, had an exposure period of 3454 s, and was taken on 1996/09/12. By contrast, the LWS observations were centred on $\alpha$ = 00h 13 m 00.94 s, $\delta$ = 72° 31' 19.8", were taken as part of program TDT47300616, had an exposure period of 1318 s, and took place on 1997/3/3.

## 3. The Spectral Properties of NGC 40 in the NIR and MIR

### 3.1 The Properties of the Emission Spectrum, and Implications for Abundances

The ISO spectrum in Fig. 1 reveals that NGC 40 is associated with several differing longer-wave emission components. It is possible to identify a broad continuum, multiple ionic transitions, and a variety of narrow emission bands, including an excess between ~ 27 and 45 μm (the so-called 30 μm feature), and an emission plateau between ~ 19 and 24 μm (the 21 μm feature). The broader component is modelled using a function $\varepsilon(\lambda)B(\lambda,T_{GR})$, where $B(\lambda,T_{GR})$ is the Planck function for grain temperatures $T_{GR}$, and $\varepsilon(\lambda) \propto \lambda^{-\beta}$ is grain emissivity. It is apparent, when fitting this to the MIR and FIR spectra (the combined LWS+SWS spectrum in indicated in the insert to Fig. 1), that the best representations of this data are obtained where $\beta$ is ~0→1, and temperatures are in the region of $T_{GR}$ = 150 K ($\beta$ = 0) and 104 K ($\beta$ = 1). The $\beta$ = 2 trends can be separately fitted to the short ($\lambda$ < 40 μm) and longer wave data, although it is difficult to obtain adequate fits for both of these regimes combined – where the fit is reasonable for the FIR results, then it is poor in the MIR+NIR.

Some care must be taken in the interpretation of these trends, however. In fitting the LWS and SWS data at wavelengths close to ~44 μm, we have found it necessary to reduce LWS fluxes by a factor of 2.8. This suggests that the emission is substantially extended with respect to the SWS aperture (20x 22 arcsec$^2$), and that it may even extend beyond the limits of the LWS aperture (~70 arcsec; see Sect. 2). It is therefore clear that the strength of the continuum may be being significantly underestimated. Where, in addition, the size of the emission regime changes with wavelength, such that (say) the source appears larger at larger values of λ, then it's clear that the combined LWS +SWS spectrum may be appreciably distorted.

Finally, we have assumed that the underlying continuum can be fitted using single temperature functions alone. Where there is a gradient in temperatures, and/or more than one temperature component, then it is clear that the fits in Fig. 1 can only be broadly indicative. The main interest of this analysis concerns the narrower emission components, however, and we shall consider what these might say about the chemical composition of the source, and previous estimates of abundance.

The 30 μm and 21 μm features have previously been noted by Hony et al. (2001) and may arise, in the case of the 21 μm feature, due to large PAH clusters, hydrogenated amorphous carbon grains, hydrogenated fullerenes, nanodiamonds, SiC+SiO$_2$ grains, TiC nanoclusters and so forth (see e.g. von Helden et al. 2000; Zhang, Jiang & Li 2009, and references therein). The range of possibilities appears to be extraordinarily large, although Zhang, Jiang & Li (2009) point to nano-FeO as the most likely candidate for this emission. The main carrier for the 30 μm feature, by contrast, is likely to be MgS, or some other still unidentified carbonaceous material (see e.g. Hony et al. 2002). It has also been suggested that it may arise due to peripheral OH groups attached to aromatic clusters (Papoular 2000; Volk et al. 2002).

The 30 μm feature has been observed in several types of evolved C-rich sources, ranging from AGB stars to PNe (Forrest et al. 1981),

whilst the 21 μm plateau has also been observed in a narrow range of post-AGB sources which are C-rich, metal poor and s-process enhanced (Kwok et al. 1999; Van Winckel & Reyniers 2000).

A PAH component at λ11.3 μm is also evident in a Spitzer spectrum of the source (see the superimposed Spitzer spectrum in Fig. 2, and our further comments in Sect. 2). The Spitzer results do not however confirm the plateau between ~12.4 μm and 14.5 μm, which is detected in ISO, and may be attributable to PAH emission features yet again. This may indicate that the ISO "plateau" is an instrumental artefact, or more likely, that the differences result from the differing sizes of instrumental aperture. The Spitzer aperture has an area which is ~7 times smaller than that of ISO, and a quite differing projected shape (see Sect. 2).

Several other features in this spectrum may arise from instrumental artefacts (see e.g. Price, Sloan & Kraemer 2002), correspond to noise, or represent still unidentified emission components. The otherwise unidentified 14.6 μm feature in Fig. 2 corresponds in wavelength to α-$Si_3N_4$, for instance (see e.g. Pitman, Speck & Hofmeister 2006), although the absence of other (sometimes stronger) nitride features argues against this identification.

It is therefore clear that the grains in NGC 40 are responsible for a complex MIR spectrum, and suggest C/O ratios which are very much in excess of unity. This is consistent with the CEL abundances of a variety of authors (Clegg et al. 1984; Aller & Czyzak 1979; Pottasch et al. 2003), but inconsistent with those determined using ORLs (Liu et al. 2004). It is also consistent with the detection of fluorine in this source by Zhang & Liu (2005), a rather unusual finding when one considers Galactic PNe taken as a whole. Zhang & Liu point out that $^{19}F$ is most abundant where C/O ratios are large – and that it may be particularly abundant in NGC 40 because of the high density/mass-loss CS outflow.

### 3.2 Implications of the Spitzer/ISO Spectra for IRAC imaging of NGC 40

The MIR images of NGC 40 to be discussed in Sect. 4 may arise as a result of a variety of solid state and gaseous emission processes, including warm dust continuum emission; polycyclic aromatic hydrocarbons (PAHs); lower excitation ionic transitions, and shock or fluorescently excited molecular hydrogen. The ISO and Spitzer spectra of the source, illustrated in Figs. 1 & 2, enable us to gain an overall perspective of what may be important in this respect.

We note for instance that the 8.0 $\mu$m IRAC band is dominated by PAH emission features (see Fig. 2; where the individual IRAC bands are indicated using differingly coloured profiles), although there is little evidence for the 3.3 $\mu$m PAH feature. This component is normally quite weak in PNe, however, and the S/N of these results is rather poor. The $\lambda$6.985 $\mu$m and $\lambda$8.991 $\mu$m transitions of [ArII] and [ArIII] are also important within the 8.0 $\mu$m band, whilst the absence of lines such as [ArV] and [NeV], and the relative strengths of [NeII] and [NeIII] confirms the low excitation nature of the shell at MIR wavelengths. We also note that emission at 8.0 $\mu$m includes the Wien limits of the broader grain continuum, whilst there is evidence for significant Br$\alpha$ emission within the 4.5 $\mu$m band. Pfund transitions occur in the 3.6, 4.5 and 8.0 $\mu$m channels.

Despite the limits upon the quality of the shorter wave observations, we note that $H_2$ S(1), S(2), and Q branch transitions have been detected in the 2.1-2.4 $\mu$m regime, corresponding to excitation temperatures of $\sim$ 2100 K (Hora, Latter & Deutsch 1999) – a value which suggests shock excitation of the molecular envelope. The v = 0-0 transitions of $H_2$ might therefore be expected to contribute to all of the IRAC bands, with the S(4) to S(7) transitions being strong within the longer wave channels (5.8 and 8.0 $\mu$m).

Finally, we note that small PAH molecules are also prone to stochastic heating as a result of the absorption of individual CS photons, and that these may lead to higher temperature continua capable of being detected in the MIR (see e.g. Draine 2003). It would seem that this component is weak in the centre of this source, at least compared to the dominant fluxes at longer MIR wavelengths (Figs. 1). We see for instance no evidence for such emission in Fig. 2, although it's clear that the S/N of this spectrum leaves much to be desired. On the other hand, there is evidence that such continua may be important in the halo of NGC 40 (Phillips & Ramos-Larios 2006; Willner, Becklin &

Visvanathan 1972), and could conceivably contribute to the emission noted in our analysis in Sects. 5 & 6.

It is therefore clear that a variety of mechanisms may contribute to the MIR properties of NGC 40, and appreciably affect the morphology, fluxes and profiles discussed below.

## 4. IRAC Imaging of NGC 40

An IRAC image of NGC 40 is illustrated in Fig. 3, where we show combined unsharp masked results at 3.6 μm (green), 4.5 μm (red) and 5.8 μm (blue). The 8.0 μm results have been excluded since they show appreciable contamination by a central emission band. They will however make an important contribution to the analysis of emission mechanisms within this source, and to our discussion of the structures of the halo and rings.

Several features are apparent from a casual inspection of this image, including the bright inner elliptical shell (with dimensions ~ 41x48 arcsec$^2$); several rings within the inner halo (the NE components of which are labelled 1 through 4; see further discussion of these in Sect. 5). We also note the presence of several radial emission spokes outside of the inner elliptical shell; the presence of an irregular band of emission outside of the NE limits of the halo (labelled A); and a jet-like feature (labelled B).

Contour maps of NGC 40 are shown in Fig. 4, where we again see a marked change in morphology with radial distance from the nucleus – a variation from high levels of ellipticity at the centre, to approximate circularity in the halo. There is also a bright radial bar outside of the northeast limits of the inner halo – a feature which has been previously identified as a jet (labelled B in Fig. 3; see e.g. Martin et al. 2002; Meaburn et al. 1996). Such an hypothesis also gains credence from the presence of a similar (but much smaller) feature on the opposing side of the shell (not visible here, but clearly evident in visual images of the source). A problem with this hypothesis is that the kinematics of the bar appear untypical of jets (Meaburn et al. 1996), whilst there is little evidence that the "jet" extends into the inner portions of the shell. Similarly, we note that the outer regions of the envelope appear to be associated with many smaller tongue-like features, interpreted by Martin et al. (2002) as Rayleigh-Taylor instabilities. The so-called counter-jet on the SW side of the shell may simply represent an enhanced version of one of these tongues.

Further evidence that we are unlikely to be dealing with a genuine jet outflow may also be discerned in the MIR image in Fig. 3, from which it is apparent that there are many bright emission spokes extending deep into the halo, presumably arising from the escape of UV photons through the clumpy interior shell. These also lead to some irregularity of the ring-like structures. It is therefore conceivable that the larger jet-like feature (and smaller counter jet in the visible) are little more than brighter examples of the radial emission spokes. We shall discuss the possible reasons why this spoke is more enhanced in our further analysis below.

It is worth remarking that such spoke-like structures are by no means unique, and similar structures have been observed in IC 2165, NGC 2867, NGC 6543, and NGC 6772 (Corradi et al. 2003; Balick 2004; Ramos-Larios & Phillips 2009). It is also clear that they must complicate any analysis of emission line strengths and nebular abundances. On the other hand, such variable levels of excitation may also cause strong spatial variations in temperature $T_e$, and contribute to differences in the ORL and CEL abundances noted in Sect. 1.

We finally note that the clumpy (and partial) ring of emission to the NE side of the shell (labelled A in Fig. 3) has been interpreted as being the result of shock compression by the ISM (Martin et al. 2002); an hypothesis which is consistent with the asymmetrical pattern of outer filamentary emission (see Sect. 1). Although the HIPPARCOS proper motion of the central star appears to imply motion to the north and west (viz. Martin et al. 2002; Perryman et al. 1997), the errors on these measurements are extremely high, and accuracies are of order ~$1\sigma$ for $\mu_\delta$, and ~$1.7\sigma$ for $\mu_\alpha$. Martin et al. (2002) suggest that the partial annulus of emission may represent a distorted segment of the "jet", which has been bent through contact with the ISM. Such an hypothesis is by no means unique, however, and is not particularly consistent with Fig. 3. We regard it as more likely that the band is a result of shocking between the outer halo and the ISM, leading to locally enhanced compression and densities. The "jet" may therefore represent nothing more than an enhanced emission spoke within this higher density shock compressed gas.

Finally, we illustrate the 3.6μm/4.5μm, 5.8μm/4.5μm and 8.0μm/4.5μm ratio maps in Fig. 4. Since there is very little PAH emission within the 4.5 μm photometric channel, such ratios tend to show the distribution of PAHs where these are dominant within a source. The effect of the saturation of the central star in the 8.0 μm band is clearly evident in the centre of the source, where it leads to a dark flattened "scar" (length ~20 arcsec) extending to the east. The saturation also leads to a smaller circular region of lower ratios to the right of the central star (i.e. to the west). Apart from these small central distortions, the effect of saturation appears to be minimal.

Several characteristics are apparent from an inspection of these maps. The first is that the radial emission spokes are also evident in the 5.8μm/4.5μm and 8.0μm/4.5μm ratio maps, where they lead to lower ratios compared to the immediately surrounding halo regime. This could arise where higher intensities of radiation field cause sublimation of smaller PAH molecules, resulting in the levels of PAH emission being correspondingly reduced. By contrast, the 3.3 μm feature appears to be relatively weak (Fig. 2), and this may explain why the spokes are less apparent in the 3.6μm/4.5μm ratio map. This may arise because whilst the 3.3 μm PAH emission is due to the aromatic C-H stretching mode, those at 6.2 and 7.7 μm are due (in part) to the C-C stretching mode. Ionisation of the PAHs (and a reduction in the numbers of C-H bonds) may very well lead to a reduction in 3.3 μm emission compared to longer wave PAH emission bands.

It would also appear that the ratios decrease in the outer "jet" component, suggesting that PAH emission in this region may be somewhat reduced. By comparison, the 5.8μm/4.5μm and 8.0μm/4.5μm ratios appear to be larger in the nucleus, decrease in the annular regime between r ~ 20 and 35 arcsec; and increase yet again towards outer portions of the halo. This trend will also be confirmed in profiles taken through the centre of the source (see Sect. 5). Similar variations are noted in the 3.6μm/4.5μm map, although they appear again to be less extreme than for the longer wave ratio results. Such changes may arise because of the stronger prevalence of emission spokes within this regime, and the effect that they have suppressing PAH emission bands.

## 5. MIR Profiles through NGC 40, and the nature of the Rings and Halo

Profiles through the centre of NGC 40 are illustrated in Fig. 5, where we show radial slices through the centre of the source, and for all four of the photometric bands. Note that although the 8.0 $\mu$m results are affected by a contaminating emission band, extending E-W, and caused by saturation of central star (see Sect. 2), this has little effect upon these particular results. The bright elliptical shell extends between relative positions (RPs) ~ ± 26 arcsec, whilst the outer halo commences at distances RP > 28 arcsec. The SW portion of the profile is strongly affected by a weak field star, whilst the NW sector shows evidence for at least four annular rings, and the partial outer emission band noted in Figs. 3 & 4. The individual SW and NE ring features are indicated by arrows (and labeled SWR1, SWR2, NER1, NER2 & etc.), whilst the band of northeasterly emission is flagged as being an ISM/Nebular interface, consistent with our discussion in Sect. 4.

It is clear that all of the profiles fall-off in a similar manner, and have almost identical variations in surface brightness – a trend which is at variance with what is observed in many other PNe (see e.g. Phillips & Ramos-Larios 2008, 2010; Ramos-Larios & Phillips 2008; Ramos-Larios, Phillips & Cuesta 2008). This leads to the approximate invariance in surface brightness ratios noted in Fig. 6. It is also clear that the rings are responsible for small deviations in the profiles, whilst the ISM/Nebular interface is dominant where RP > 80 arcsec.

We have attempted to better define the properties of the rings and outer interface by representing the underlying, smoother halo emission by a series of fourth-order polynomial splines, least-squares fitted to the observed surface brightness fall-off. This component is then removed from the total halo emission to yield the fluxes of the ring and interface components.

The results are illustrated in Fig. 7 (upper panel), where it is clear that 5.8 and 8.0 $\mu$m fluxes are the most dominant, and the 3.6 and 4.5 $\mu$m fluxes are reasonably comparable. These rings are also labelled in Fig.

3, where it will be seen that ring NER1 is actually quite complex; it consists (along this particular PA = 55°) of two closely spaced components – a fainter outer ring feature and much brighter inner component. These are not resolved in the profiles in Fig. 7, which tend to be dominated by the strongest of the pair. The separation between the rings increases towards larger values of PA. A further interesting representation of ring and interface trends may be obtained where one divides their fluxes by the total halo emission (see Fig.7, lower panel). It follows, in this latter case, that where levels of fractional emission are of order < 0.2, then they indicate the approximate ratio of ring-to-halo surface brightness. Where the ratio approaches of order unity, on the other hand, then levels of underlying halo emission are relatively small, and the superimposed feature is responsible for most of the emission.

It is clear that the brightness of the rings, taken as a whole, is of order ~ 20-30% of that associated with the halo; a value which is rather large when compared to the optical ([NII] + H$\alpha$) estimates of Corradi et al. (2004) (~10%). It would also seem that the rings are proportionately stronger at 3.6 and 4.5 $\mu$m, and weaker at 5.8 and 8.0 $\mu$m – the reverse of what might have been expected from the trends in Fig. 7 (upper panel), but similar to what is observed in several other PNe (Phillips et al. 2009; Phillips & Ramos-Larios 2010). Comparable trends are also noted for the regime of ISM/Nebular interaction, although it is clear that fractional levels of emission for this feature are very much larger.

It is finally worth noting that the peak halo surface brightness in Fig. 7 (upper panel) is of order R(MIR) ~1/12 of the maximum value within the core. Although there are no comparable profiles for this source in the visible, Corradi et al. (2003) note that most such halos have a value R(VIS) < 4 $10^{-3}$ – whilst their sample as a whole indicates ratios < 2.5 $10^{-2}$. The halo/core surface brightness ratio in the MIR is therefore likely to be much greater than is observed in the visible.

This disparity has previously been noted for other PNe (Phillips et al. 2009; Phillips & Ramos-Larios 2010), and is the reason why the halo and rings are visible in the MIR – even though instrumental sensitivities may be less than those applying in the visible. We shall now consider what these results may portend for emission processes within these regimes.

## 6. The Nature of Emission in the Halo, Rings, and ISM/Nebular Interface

We have used the results in Figs. 5 & 7 to determine flux ratios 3.6μm/5.8μm and 4.5μm/8.0μm for the envelope of NGC 40, including the rings, halo, and outer ISM/Nebular interface. The results are indicated in the diagnostic diagram in Fig. 8, based upon results taken from Reach et al. (2006), and the trends expected for smooth grain continua.

The individual regions indicated in this figure correspond to the colours for PAH band emission; the range expected for molecular transitions (primarily $H_2$); and emission associated with ionized gas (IONIC). Although this figure was originally devised to investigate MIR emission in Galactic supernovae remnants, and should not be used for PNe without some degree of qualification, we believe it is of utility for the present analysis. Thus, the range of IONIC ratios takes account of the shock emission expected for H, $Fe^+$, $Ni^+$, $Ar^+$ and $Ar^{++}$. Although the emission from PNe is normally dominated by unshocked (photoionised) emission, this is not necessary the case for the rings of these sources, given that there is evidence that shocks may be important in generating these features (see e.g. Hyung et al. 2001; Phillips, Cuesta & Ramos-Larios 2010; and the shock modelling summarised in Phillips et al. 2009). Similarly, the interaction of halos with the ISM, observed in NGC 40 and other PNe, would be likely to lead to appreciable levels of shock induced emission. Evidence for (lower excitation) shock enhanced transitions has been noted in M2-9, NGC 6905 and NGC 7009, among other sources (see e.g. Phillips & Cuesta 1999; Cuesta, Phillips & Mampaso 1993; Phillips, Cuesta & Ramos-Larios 2010). Where this is not the case, then an analysis of case B emission for pure hydrogen plasmas having $T_e = 10^4$ K, and including the Pfα (7.46 μm), Pfβ (4.65 μm), Pfγ (3.74 μm) and Brα (4.05 μm) transitions, would imply relative intensities (0.25/3.7/0/1) in IRAC bands (1,2,3,4), and ratios $\log(I(4.5\mu m)/I(8.0\mu m)) \sim 0.56$ and $\log(I(3.6\mu m)/I(5.8\mu m)) \to \infty$ - values which are well outside of the range observed in our present results. Similar caveats apply to the analysis of $H_2$ emission, since the region of molecular emission in Fig. 8 refers primarily to shocked transitions. Given the probable shock activity in the halo and rings of NGC 40, and the evidence for shocked $H_2$ in many other PNe as well (e.g. Hora, Latter & Deutsch 1999; Sahai et al. 1998; Bujarrabal et al. 1998; Rudy et al. 2001; Hora & Latter 1994; Ramos-Larios, Guerrero & Miranda 2008), we believe that this regime is of most relevance for the present analysis. It may not however be applicable where fluorescent excitation occurs.

Finally, we remark that the PAH emission regime defined by Reach et al. was based (primarily) upon an analysis of the reflection nebula NGC 7023, the emission from which was presumed to be dominated by PAH emission bands. In reality, no PAH emission bands have been detected in the 4.5 μm IRAC channel, and the pure PAH emission regime may require shifting to the left.

Taken all-in-all, therefore, we believe that such a diagram is useful for analysing the present source, and it will be so employed here. Care must be taken, however, when applying it to other PNe.

Finally, dust continua for differing exponents $\beta$ and grain temperatures $T_{GR}$ are indicated by the diagonal lines to the lower right-hand side. The grain continua are assumed to vary as $\propto B_\nu(T_{GR})\varepsilon(\nu)$, where grain emissivity $\varepsilon \propto \nu^\beta$, and the results take account of filter throughputs over the relevant IRAC passbands.

Uncertainties in these results may arise from a variety of causes, and lead to systematic and random errors. It is clear for instance that fluxes for the rings are particularly low in the shorter wave (3.6 and 4.5 μm) channels, and that this may lead to random errors of < 20% in the estimation of ratios. This maximum error uncertainty is indicated in Fig. 8. Apart from this, the use of differing fit parameters for the underling halo emission may also cause systematic movement of points within the colour plane – although we believe that the effects of this uncertainty are small.

Fluctuations in halo emission, over and above those due to the rings themselves, may also lead to small systematic shifts in the colour plane, whilst weak background stars have a similar effect, and represent a further (and more insidious) contribution. The effect of one such star is clearly visible to the SW side of Fig. 5, where it causes massive distortion in the profile close to the rings SWR2 and SWR3. This contribution at least has the benefit of being clearly apparent and avoidable. Other weaker stellar contaminants may be less easy to recognise and remove.

Such systematic trends are, for the most part, of relatively little importance to Fig. 8 – and indeed, we have selected this precise profile direction so as to minimise these contributions. We believe that the general tendency of the results is trustworthy, and is for the most part consistent. However, it is possible that the 3.6 μm flux for NER2 is somewhat too low, shifting the location of this source to lower positions within Fig. 8, whilst ring NER3 was not detected at 4.5 μm, and is therefore excluded from the colour plane.

It is clear that most of the outer emission components of NGC 40 fall with the upper right-hand quadrant of Fig. 8, and are well outside of the regime of PAH band emission. This is rather unexpected given the spectrum in Fig. 2, but does not of course imply that PAH emission is excluded as a whole. It does however suggest that other mechanisms may be more (or equally) important – and in particular, that much of the emission may arise from shock excited $H_2$. The halo may, if this is the case, be similar to those noted in NGC 2440 (Latter & Hora 1997; Wang et al. 2008) NGC 6881 (Ramos-Larios, Guerrero & Miranda, 2008), NGC 6853 (Kwok et al. 2008) and other nebulae, where extended NIR emission is attributed to the S(1) v = 1-0 transition of $H_2$.

Although this appears to be the most straightforward interpretation of the MIR ratios, and consistent with the detection of $H_2$ in the NIR (see Sect. 3), it is also possible that we are dealing with a combination of two or more differing emission mechanisms. It is conceivable for instance that a mixture of warm dust continua, together with PAH emission bands might swing the relevant data points into the $H_2$ regime. Values of $T_{GR}$ for such a situation would be expected to range between ≈ 500 K (for β = 2) and 620 K (β = 1). Given that such large temperatures normally require small grain sizes (see e.g. Phillips & Ramos-Larios 2008), exponents β = 0 can probably be excluded for this case.

Where this combined warm dust + PAH emission is indeed responsible for the emission, then it is clear that the proportion of PAH emission in the 8.0 μm channel would be required to be reasonably high – of order ~90 % for the halo as a whole, and ~65% for the outer interaction regime. The contribution of the PAH bands in the shorter wave channels would be significantly less.

Finally, we note that the only exception to these trends appears to be the ring NER2, for which ratios are consistent with dust continuum and/or ionized emission. The lower placement of this point seems rather peculiar, however, and is inconsistent with all of the other trends. It is therefore possible, as noted above, that fluxes for this ring are unreliable.

## 7. Conclusions

We have presented MIR imaging and spectroscopy of NGC 40 based upon results deriving from ISO and Spitzer. It is shown that the MIR spectrum for the central regions of the source contains a broad range of emission components, and a dominant continuum arising from cool (~100-150 K) grains. The presence of several PAH emission bands, and the 21 $\mu$m and 30 $\mu$m emission features is consistent with the high C/O abundances deduced from the analyses of CELs; the results are not consistent with abundances derived using ORLs. It would also seem that several ionic lines ([ArII], [ArIII] and Br$\alpha$), the longer wave cool dust continuum, and various PAH emission bands contribute to the Spitzer IRAC band fluxes. Although the shorter wave ISO spectra have a low S/N, and yield little information concerning the primary emission mechanisms, it is likely that $H_2$ transitions and warm dust emission contribute to fluxes observed in the shorter wave photometric bands.

We note that combined IRAC imaging of NGC 40, when processed using unsharp masking, shows the presence of the annular rings detected in H$\alpha$+[NII]; radial spokes of emission outside of the bright nebular core; a fragmented emission fringe to the NE of the main nebular shell; and a supposed jet-like feature which extends ~ 20 arcsec beyond the inner nebular halo. It is suggested that the north-easterly fringe is likely to represent a region of interaction with the ISM, where the AGB halo is being compressed, and local densities are appreciably higher; an hypothesis which is in broad agreement with the previous analysis of Martin et al. (2002). It is also possible that the supposed jet-like feature is little more than a radial emission spoke, created by the escape of UV photons through the clumpy interior shell.

The increased brightness of this structure compared to other such features may be a consequence of the higher gas densities caused by local shock compression.

An analysis of emission within the rings and outer NE fringe suggests that whilst fluxes are greater at 8.0 and 5.8 $\mu$m, the levels of fractional emission (rings/halo) are stronger at 3.6 and 4.5 $\mu$m; a result which matches the trends observed in several other PNe. The outer NE fringe, which we identify as arising from interaction with the ISM, has the highest relative levels of MIR emission, and it is clear that it is appreciably fragmented, and very much stronger than the underlying halo.

Although it is apparent that several emission mechanisms are capable of explaining these exterior features, it seems likely that much of the flux arises from $H_2$ transitions. This would suggest that NGC 40 is similar to NGC 2440, NGC 6853 and other PNe, where extended NIR emission is attributable to shock or fluorescently excited $H_2$. Alternatively, it may be possible that the halo emission arises from a mixture of differing mechanisms, including PAH band components and a warm ($T_{GR} \approx$ 500-620 K) dust continuum. Evidence for such hot dust emission has been provided by Phillips & Ramos-Larios (2006) and Willner, Becklin & Visvanathan (1972), although there is little evidence for it in the (low S/N) ISO spectrum of the central regions of the source. For this latter case, the proportion of PAH-to-total emission in the 8.0 $\mu$m channel would be required to range between ~0.65 (for the outer shock interaction regime) through to ~0.9 (for the halo as a whole).

**Acknowledgements**

We would like to thank an anonymous referee for several interesting and perceptive remarks. This work is based, in part, on observations made with the Spitzer Space Telescope, which is operated by the Jet Propulsion Laboratory, California Institute of Technology under a contract with NASA. GRL acknowledges support from CONACyT (Mexico) grant 132671 and PROMEP (Mexico).

**Figure Captions**

**Figure 1**

SWS spectrum of NGC 40 acquired using ISO pipeline data, in which we indicate the primary ionic and molecular transitions in the MIR (lower bars). We also indicate the primary ionic transitions in NGC 40; the 21 and 30 μm features, and several of the PAH molecular bands. These are superimposed upon a broad dust continuum which is modelled using differing grain emissivity exponents $\beta$, and grain temperatures $T_{GR}$. The inserted figure shows the combined SWS (blue) and LWS (purple) spectrum of the source, used to obtain best fit solutions for the single temperature grain continuum functions. The fitted curves have the same colour coding as for the main figure.

**Figure 2**

The short-wave ISO spectrum for the nuclear region of NGC 40, in which we have also superimposed results deriving from Spitzer (blue squares), and indicated the main IRAC photometric filters (differingly coloured profiles, corresponding to the 3.6, 4.5, 5.8 and 8.0 μm bands). We also show again the main PAH band features and ionic and molecular transitions. Note how the broad dust continuum at longer MIR wavelengths appears to commence very close to the 8.0 μm IRAC filter. This continuum, and the 7.7 and 8.6 μm PAH band components contribute the majority of the flux within this particular channel.

**Figure 3**

A combined 3.6 μm (green), 4.5 μm (red) and 5.8 μm (blue) IRAC colour image of the nucleus and halo of NGC 40. The image has been processed using unsharp masking so as to emphasise finer details of the nebular structure. It will be seen that the annular rings detected in the visible (Corradi et al. 2004) are also apparent in the MIR, whilst we see evidence for a north-easterly jet-like feature; an attached fringe of clumpy material (attributed to interaction of NGC 40 with the ISM); and (for the first time) multiple spokes extending throughout the inner halo. It is most likely that these represent "rays" of UV photons which are penetrating the clumpy interior shell. The labels 1 through 4 identify the rings NER1 to NER4 discussed in Sect. 6 (see also Fig. 7), whilst label

A indicates the outer fringe (identified here as a region of ISM/nebular interaction), and region B is the "jet" feature.

**Figure 4**

Contour maps of NGC 40 in the 3.6, 4.5 and 5.8 μm IRAC bands (upper panels), together with corresponding 3.6μm/4.5μm, 5.8μm/4.5μm and 8.0μm/4.5μm ratio maps (lower panels). Note the evidence for an extended circular halo about the brighter elliptical core, together with a north-easterly jet-like feature and outer fringe, both of which are strong within the 4.5 μm band. The ratio maps show differences between the outer halo and inner core, and also reveal evidence for radial structures associated with the emission spokes in Fig. 3. We have used contour coefficients (A, B, C) = (0.115, 0.0666, 0.302) for the 3.6 μm map, (0.05, 0.0147, 0.328) at 4.5 μm, and (0.8, 0.527, 0.1995) for 5.8 μm. The darker shading in the lower panels indicates higher values of ratio, whilst ratio scale bars are provided in the upper parts of the panels.

**Figure 5**

NE-SW profiles through the centre of NGC 40, where we show results for all of the Spitzer IRAC bands. The width of the slice is 6.7 arcsec; the direction of the profiles is indicated in the inserted figure; and the centre of the traverse (RP = 0 arcsec) is located at the position of the (saturated) central star. The bright elliptical shell extends between RPs of ∼ -26 and +26 arcsec, whilst the outer halo contains various rings – features which are evident as deviations in the surface brightness fall-off, and referred to as NER1, NER2 & etc. (for the NE rings), and SER1, SER2 & etc (for the SE rings). The outer NE fringe of emission, attributed to interaction between NGC 40 and the ISM, is indicated as an ISM/Nebular interface. Notice how the gradients in surface brightness are similar in all of the IRAC bands.

**Figure 6**

The variation in flux ratios for a slice through the NE sector of the nebular shell. The width of the slice is 6.7 arcsec, and the direction is indicated in the inserted figure. It is clear that the ratios are, taken as a whole, relatively invariant with positional offset, and that F(8.0μm)/F(4.5μm) > F(5.8μm)/F(4.5μm) > F(3.6μm)/F(4.5μm), although there is evidence for a decrease in ratios for RPs ~ 17→38 arcsec. Errors in the ratios are relatively modest, and comparable with the sizes of the symbols where RP < 65 arcsec.

**Figure 7**

Fluxes for the rings and ISM/Nebular interaction regime in the inner halo of NGC 40, where we have eliminated underlying components of halo surface brightness (see text for details). Notice how the innermost ring is by far the brightest, and that the rings are very much stronger at 5.8 and 8.0 μm. By contrast, the lower panel shows the ratio between the ring and total halo emission, whence it is apparent that the rings have similar overall amplitudes, and are stronger than is observed in the visible (where [NII] + Hα results imply ratios of ~ 10 %). It is also apparent that 3.6 and 4.5 μm profiles have the largest relative amplitudes, with the greatest values occurring for the ISM/Nebular interface.

**Figure 8**

The positioning of the rings, ISM/Nebular interaction regime and halo in an MIR colour-colour diagram, where we also show the regimes associated with PAHs, and molecular and ionized gaseous emission. The trends to be expected for smooth dust continua are indicated by diagonal lines to the lower right-hand side, and correspond to differing grain emissivity exponents β, and temperatures $T_{GR}$. The individual ticks on these lines are separated by $\Delta T_{GR}$ = 10 K.

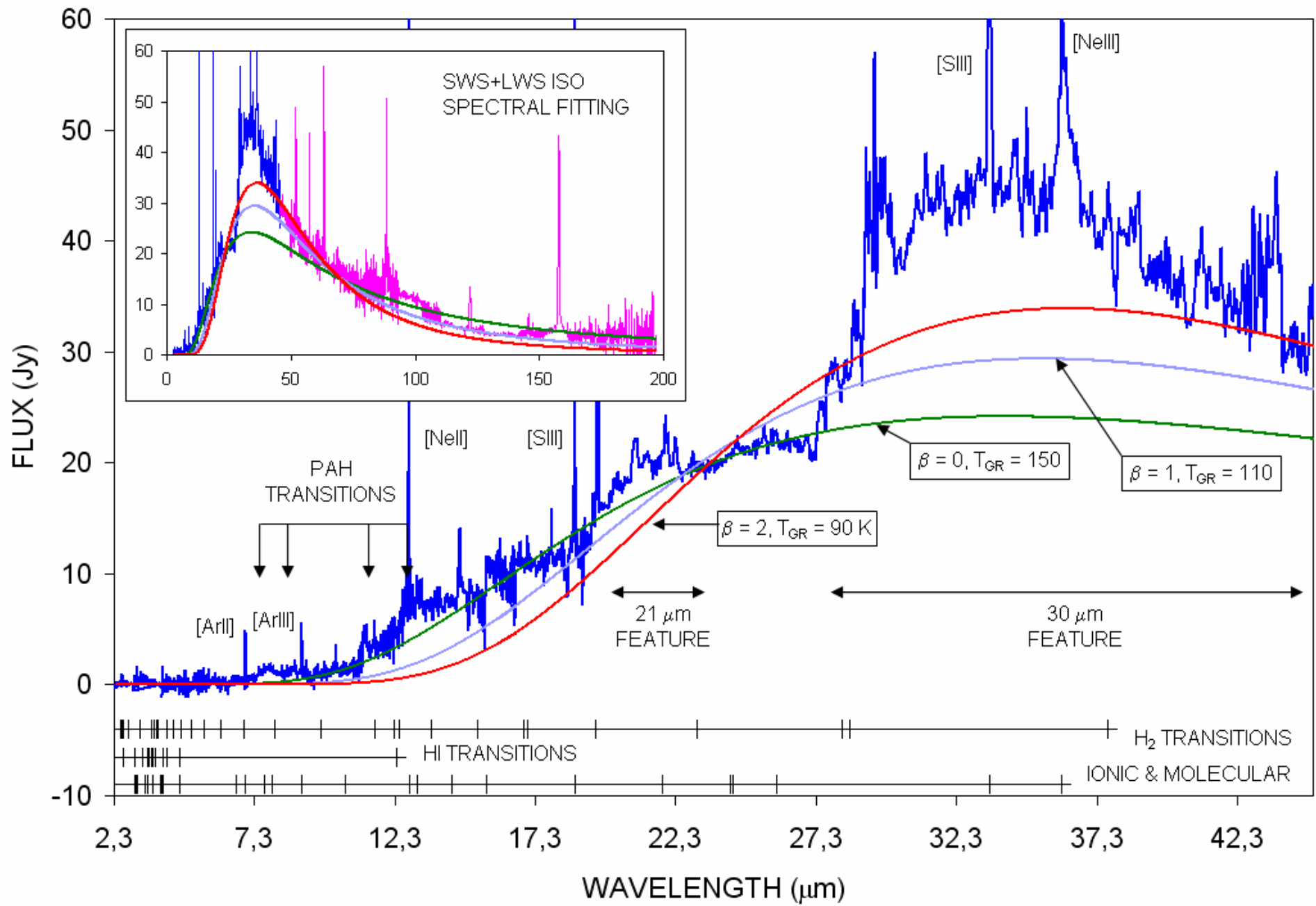

FIGURE 1

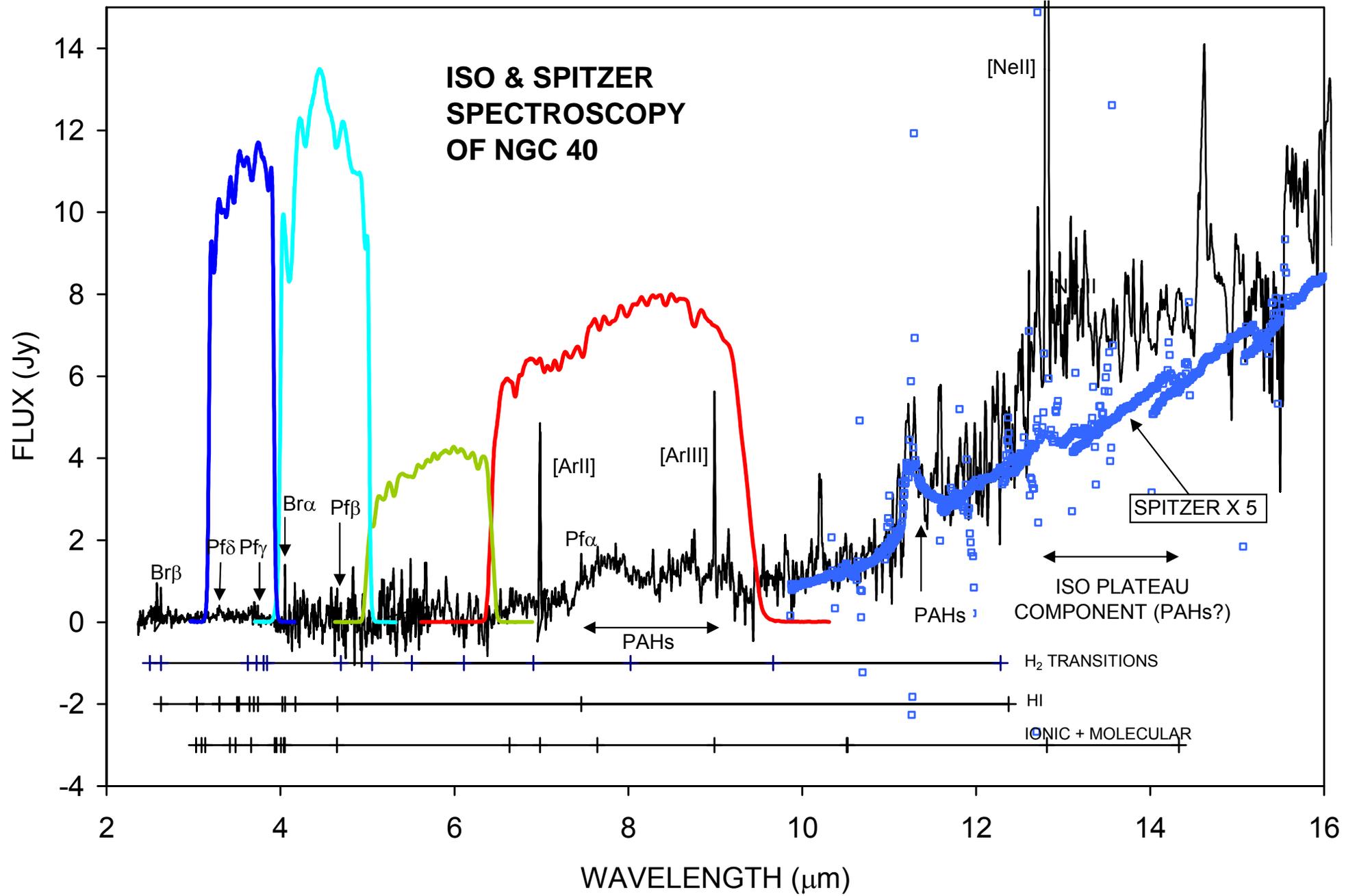

FIGURE 2

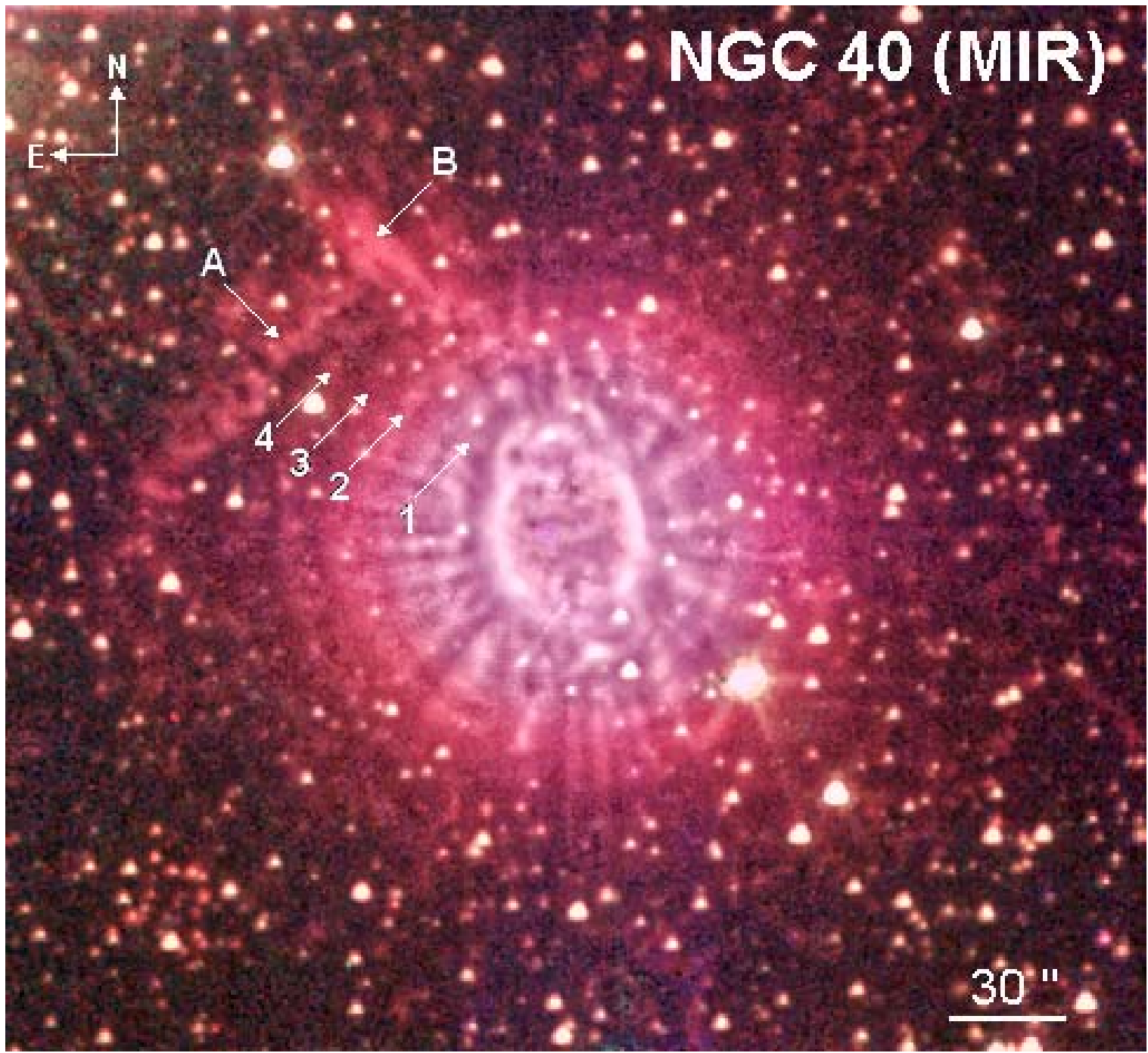

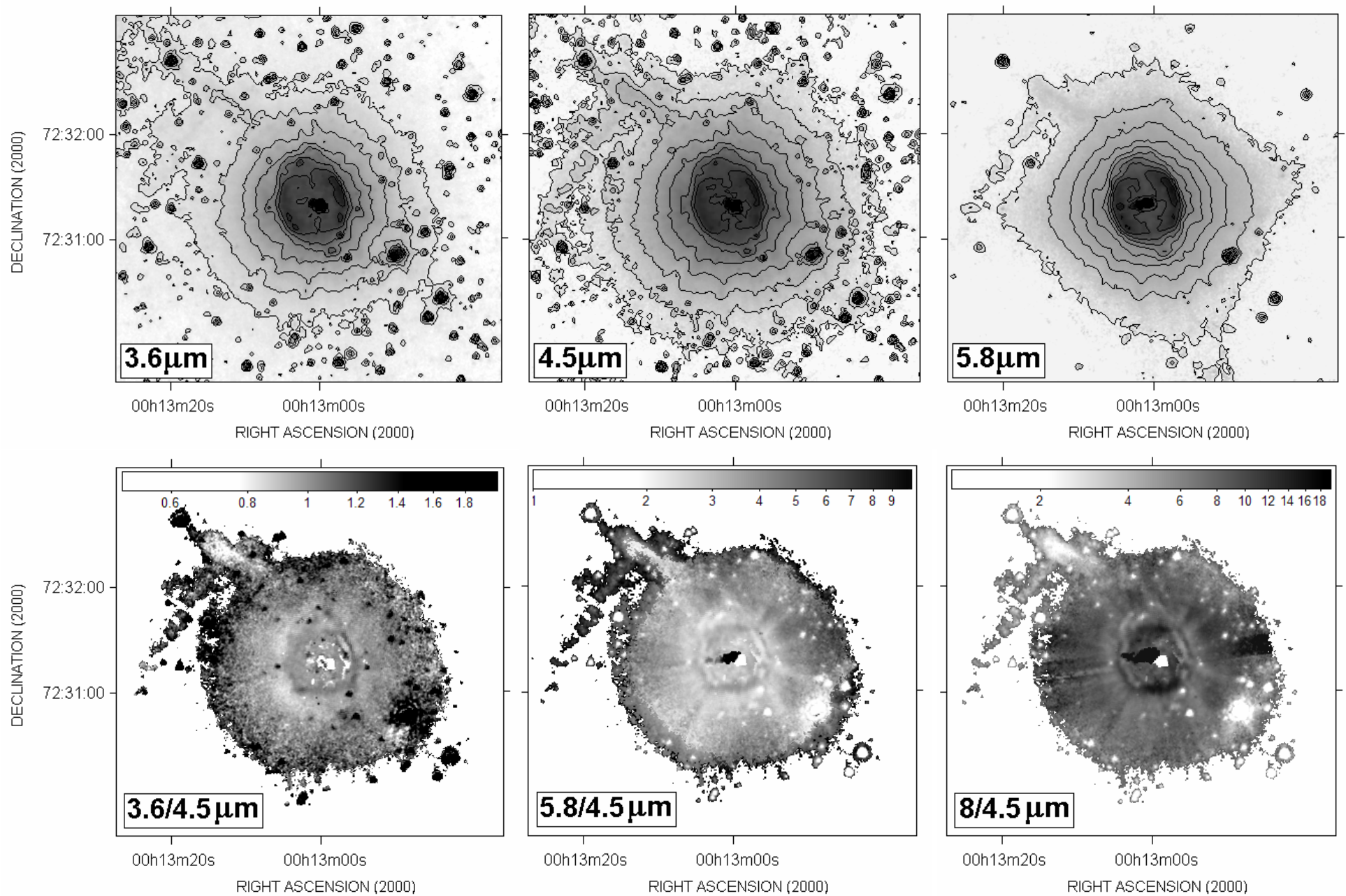

FIGURE 4

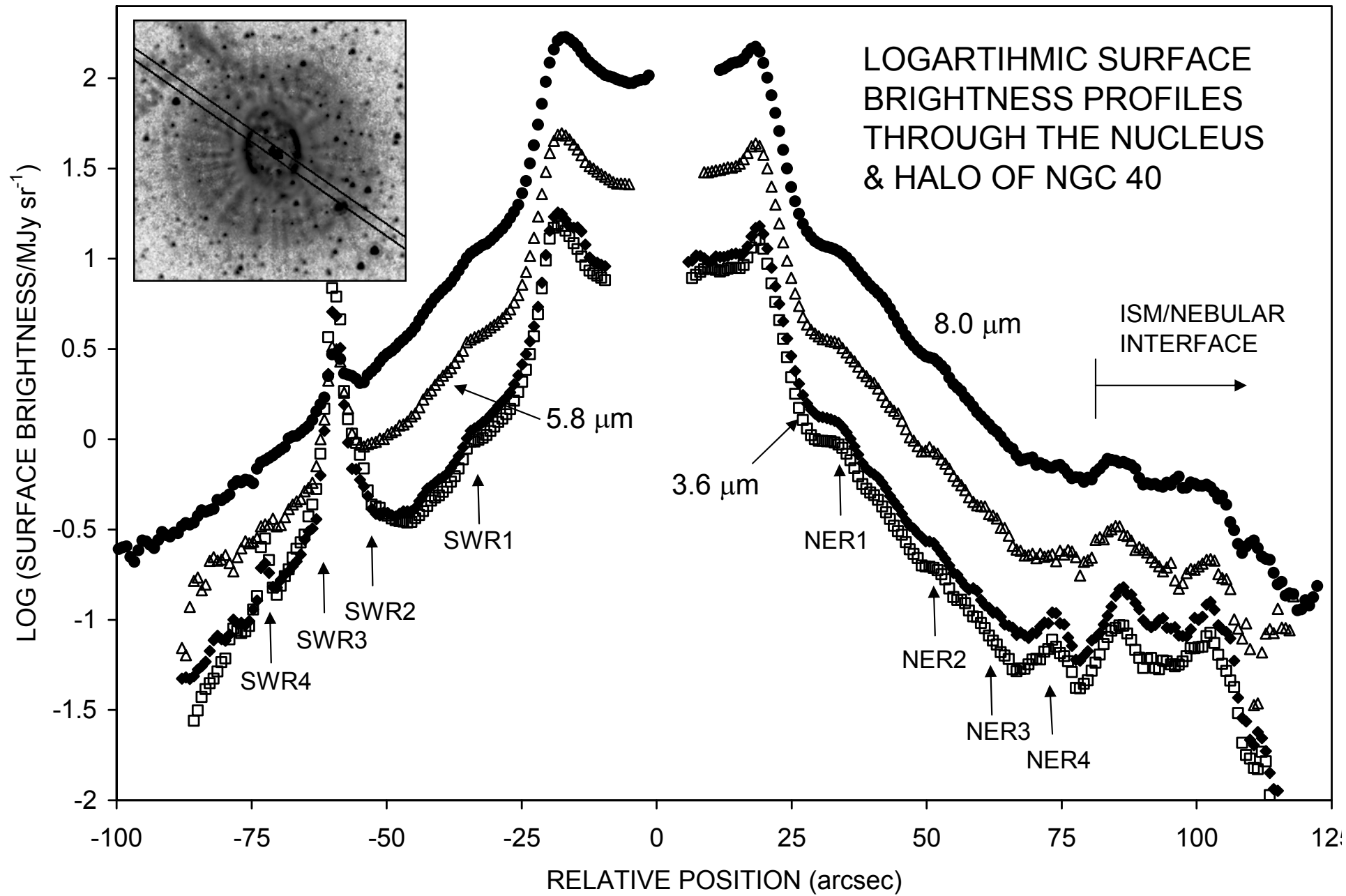

FIGURE 5

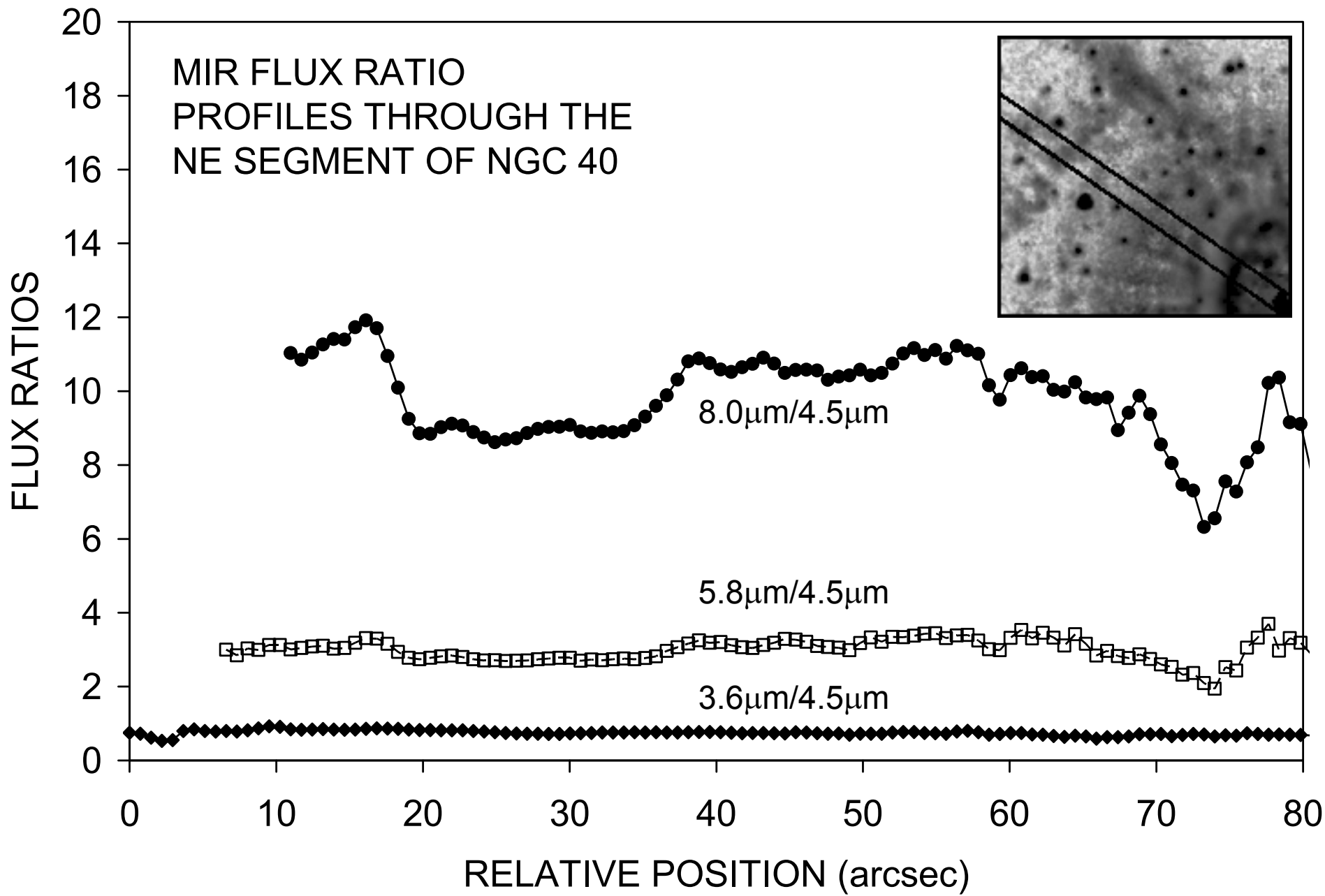

FIGURE 6

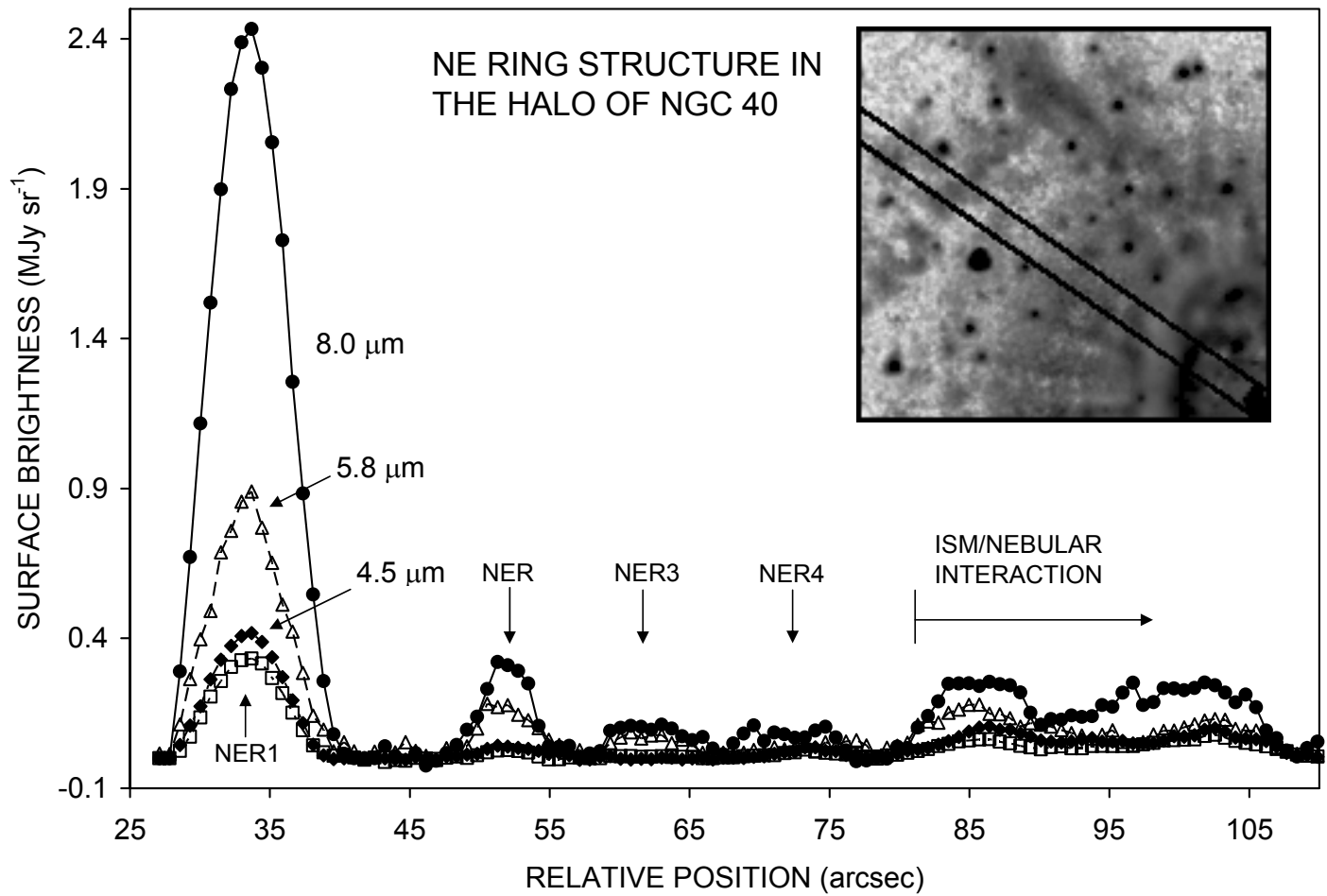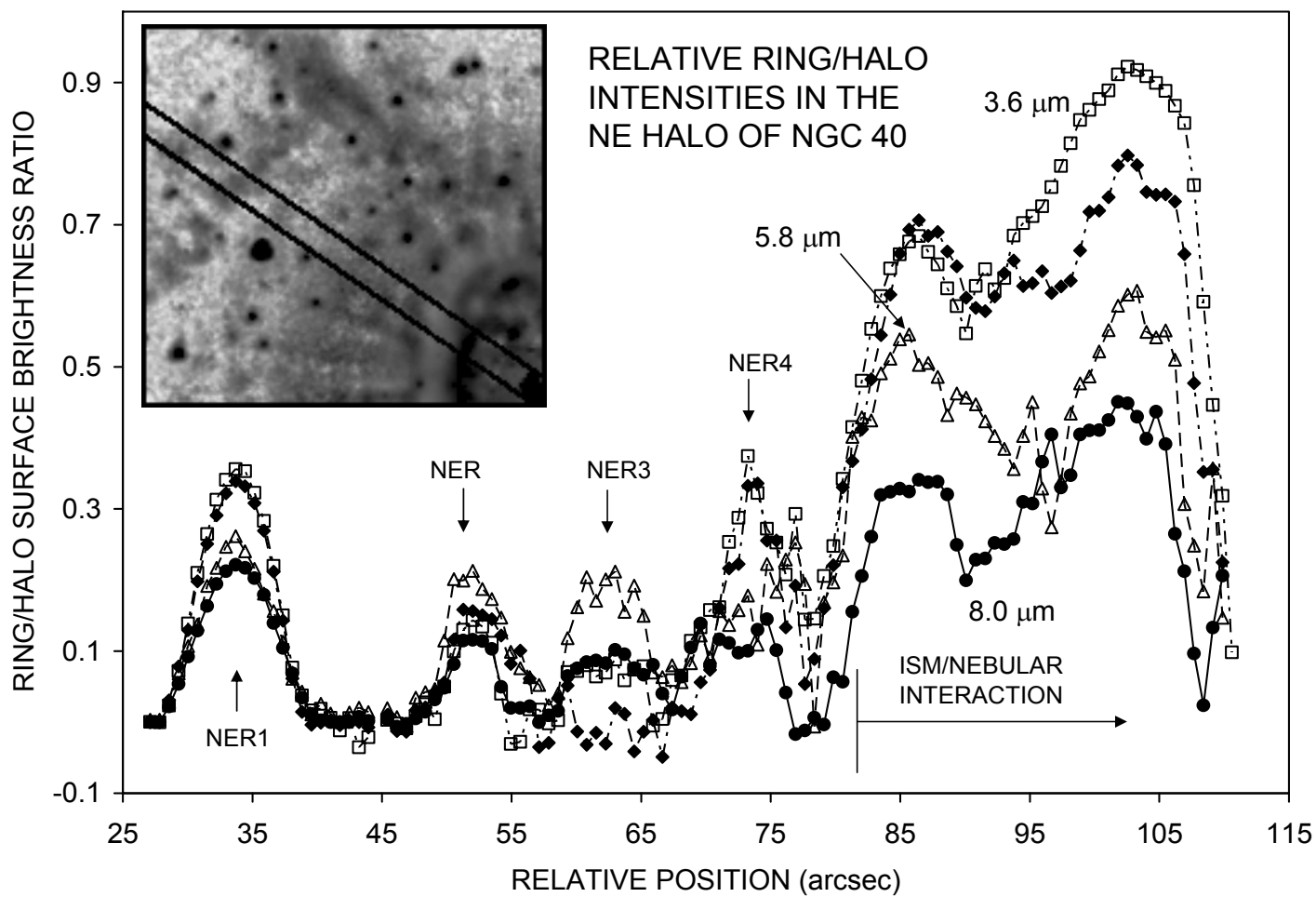

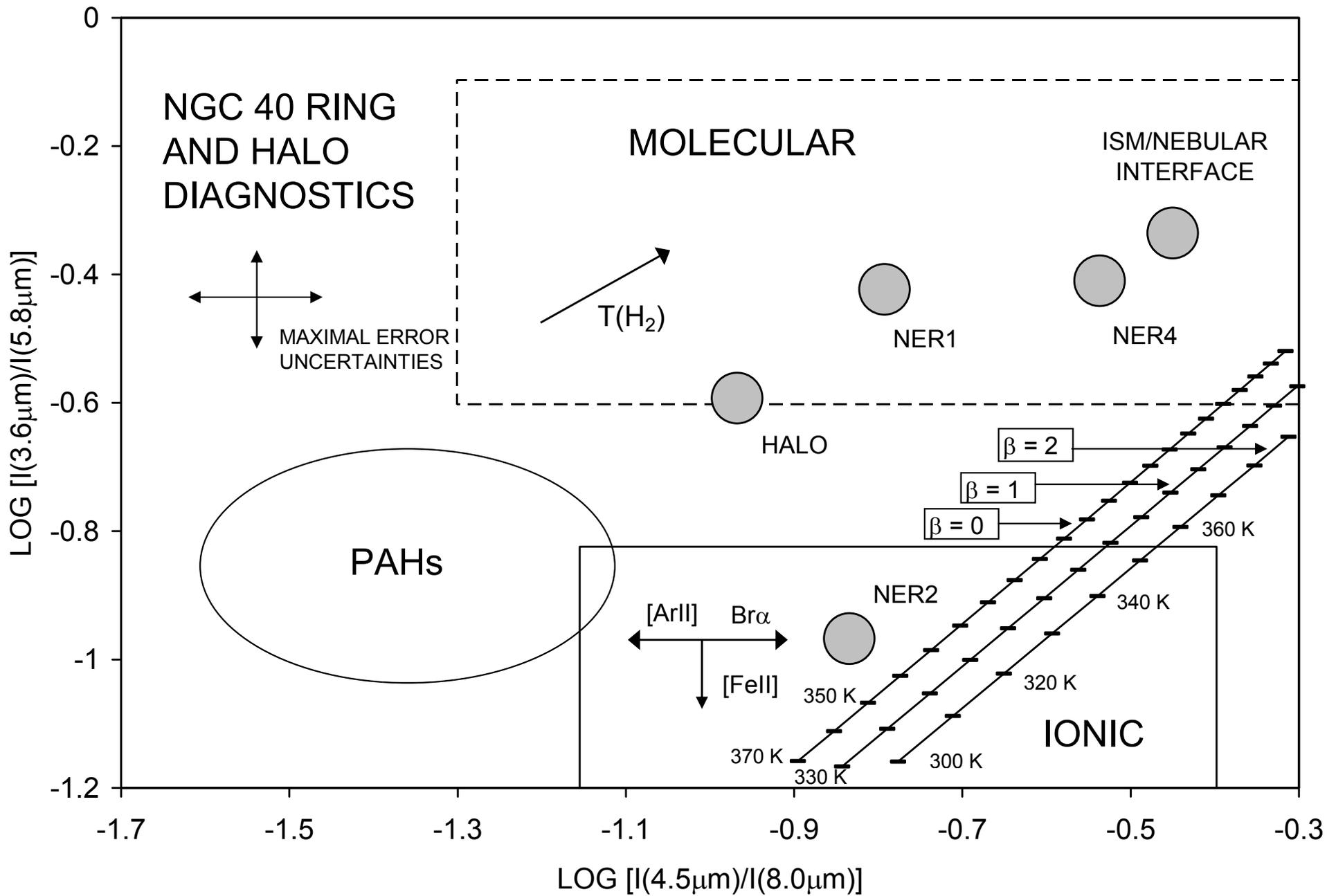

FIGURE 8